\documentclass[final]{vgtc}                          % final (conference style)
\ifpdf%                                % if we use pdflatex
  \pdfoutput=1\relax                   % create PDFs from pdfLaTeX
  \usepackage{graphicx}                % allow us to embed graphics files
  \DeclareGraphicsExtensions{.pdf,.png,.jpg,.jpeg} % for pdflatex we expect .pdf, .png, or .jpg files
\else%                                 % else we use pure latex
  \usepackage{graphicx}                % allow us to embed graphics files
  \DeclareGraphicsExtensions{.eps}     % for pure latex we expect eps files
\fi%

\graphicspath{{figures/}{pictures/}{images/}{./}} % where to search for the images

\usepackage{microtype}                 % use micro-typography (slightly more compact, better to read)
\PassOptionsToPackage{warn}{textcomp}  % to address font issues with \textrightarrow
\usepackage{textcomp}                  % use be/ter special symbols
\usepackage{mathptmx}                  % use matching math font
\usepackage{times}                     % we use Times as the main font
\usepackage{cite}                      % needed to automatically sort the references
\usepackage{tabu}                      % only used for the table example
\usepackage{booktabs}                  % only used for the table example
\usepackage{subcaption}
\usepackage{hyperref}
\usepackage{flushend}

\onlineid{1079}

\vgtccategory{Research}

\title{Inattentional Blindness for Redirected Walking Using Dynamic Foveated Rendering}

\author{Yashas Joshi%
\and Charalambos Poullis \thanks{e-mail:charalambos@poullis.org}}
\affiliation{\scriptsize Immersive and Creative Technologies Lab \\Department of Computer Science and Software Engineering\\Concordia University}

\teaser{
 \includegraphics[width=\linewidth]{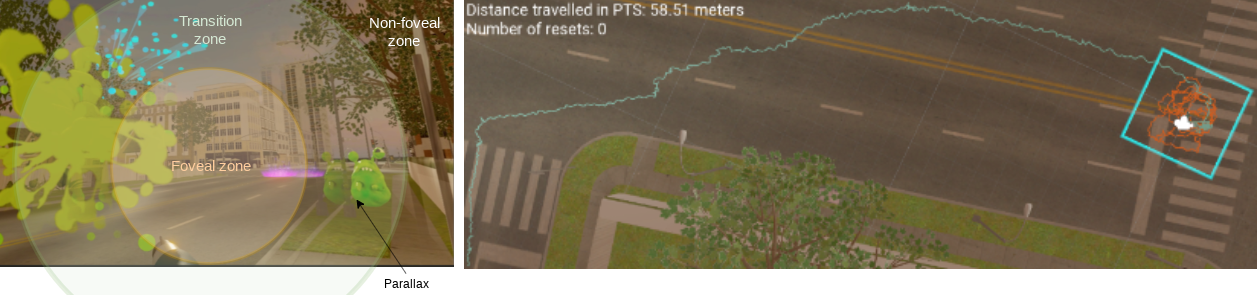}
 \caption{Left: Orientation-based redirection using dynamic-foveated rendering which leverages the effect of inattentional blindness induced by a cognitive task. The blending (e.g. parallax of green alien) in the transition zone between the rotated virtual environment (VE) (non-foveal) and the non-rotated VE (foveal) is imperceptible by the users due to inattentional blindness. Right: The path walked in physical tracked space (PTS) up to that point is shown in orange color. The corresponding path in the VE is shown in blue color. The cyan-colored box indicates the $4\times4m^2$ available PTS and the camera icon inside the box indicates the location of the user w.r.t. the PTS. Even with room-scale PTS such as the one used, the users were able to walk distances in the virtual environment (VE) which are up to almost $18\times$ orders of magnitude higher than the longest distance in the PTS; maximum recorded in experiments was 103.9m. Angular gain for the frame shown is at its maximum i.e. $13.5^{\circ}$.}
 \label{fig:teaser}
}

\abstract{%
Redirected walking is a Virtual Reality(VR) locomotion technique which enables users to navigate virtual environments (VEs) that are spatially larger than the available physical tracked space. 

In this work we present a novel technique for redirected walking in VR based on the psychological phenomenon of \textit{inattentional blindness}. Based on the user's visual fixation points we divide the user's view into zones. Spatially-varying rotations are applied according to the zone's importance and are rendered using \textit{foveated rendering}. Our technique is real-time and applicable to small and large physical spaces. Furthermore, the proposed technique does not require the use of stimulated saccades but rather takes advantage of naturally occurring saccades and blinks for a complete refresh of the framebuffer. We performed extensive testing and present the analysis of the results of three user studies conducted for the evaluation.
} % end of abstract

\CCScatlist{ 
  \CCScat{}{Computing methodologies}{Virtual reality}%
}

\begin{document}


\firstsection{Introduction}
\maketitle

Since the early days of virtual reality researchers have been investigating ways of navigating virtual environments that are spatially larger than the available physical tracked space. A number of locomotion techniques relying on pointing devices or walking in-place were proposed which have since become customary in VR applications \cite{christou2016navigation}; but as could be expected, users find these methods rather cumbersome and unnatural. The concept of redirected walking was then introduced circa 2000 \cite{1999_moshell} as a more natural way of navigating VEs, albeit with many restrictions on the shape and size of the physical and virtual spaces. 

A number of approaches have since been proposed for implementing redirected walking. Hardware-based techniques e.g. omni-directional treadmills \cite{souman2010making}, virtusphere \cite{medina2008virtusphere}, are not only an expensive solution to this problem but also fail to provide inertial force feedback equivalent to natural walking \cite{christensen2000inertial}. 

In contrast, software-based techniques are more cost effective and typically involve applying perceptually subtle rotations to the VE causing the users' to unknowingly change their walking direction. Applying these rotations to the VE, however subtle, can negatively impact the sense of immersion of the user. The reason for this is because these techniques either employ warping which introduces visual artifacts and distortions in the VE or even simulation sickness \cite{dong2017smooth}, or rely on forcing the user to look away by stimulating saccades in order to update the environment during a rapid eye movement as in \cite{sun2018towards}. 

In this work, we address the problem of redirected walking and present a novel technique based on the psychological phenomenon of inattentional blindness. Inattentional blindness refers to the inability of an individual to see a salient object in plain sight, due to lack of attention. The phenomenon was popularized in a world-famous awareness test described in \cite{simons1999gorillas} \footnote{Video of the test can be found here: \url{https://www.youtube.com/watch?v=vJG698U2Mvo}} in which participants were asked to watch a video and count the number of times a basketball is passed between the 6 people shown in the video wearing a white shirt. During the video a person dressed as a gorilla walks from right side of the screen to the left, passing through the people playing basketball, and at one point stopping to beat its chest. At the end of the experiment participants were asked to state the number of passes, and whether they noticed anything different in the video; the majority of which reported that they hadn't noticed anything out of the ordinary i.e. the gorilla walking and beating its chest.

We exploit this phenomenon and further strengthen its effect with foveated rendering. Foveated rendering is a rendering technique whose primary objective is to reduce the rendering workload. Using eye tracking the user's eyes are tracked within the VR headset in real-time. The zone in the image corresponding to the foveal vision, i.e. the zone gazed by the fovea which provides sharp and detailed vision, is then rendered at very high quality. On the other hand, the zone in the image corresponding to the peripheral vision is rendered at a much lower quality since peripheral vision lacks acuity albeit it has a wider field of view. This process is performed without causing any perceptual change to the user. Nowadays, foveated rendering is supported by hardware such as NVIDIA's RTX graphics card series which allows for real-time ray-tracing and hence, real-time performance.

Our proposed technique applies spatially-varying rotations to the VE according to the zone's importance using foveated rendering to strengthen the effect of inattentional blindness. We present the results of three user-studies. The first user study was conducted in order to determine the maximum rotation angle and field-of-view for which participants do not perceive a change. The objective of the second user study was to confirm the results of the first and also verify using only in-situ gaze redirection that a rotation of the VE results in an equivalent rotation of the participant in the PTS. Lastly, the third user study was conducted to evaluate redirected walking using dynamic foveated rendering during inattentional blindness in a small PTS. 
% Furthermore, since rotations are applied only on the non-foveal zone (the area in the rendered image corresponding to the peripheral vision), we demonstrate how two framebuffers corresponding to the foveal and peripheral zones are used to control the refresh in real-time and directly on the visible framebuffer without the user noticing. Finally, we present a user-study conducted for the evaluation of the proposed technique for redirected walking in a constrained physical space.

\noindent
The paper is organized as follows: Section \ref{sec:related_work} summarizes the state-of-the-art in the areas of redirected walking and foveated rendering. In Section \ref{sec:system_overview} we provide a technical overview of the proposed technique. The user-study for determining the maximum rotation angle and the field-of-view of the foveal zone for which change is imperceptible to a user is presented in Section \ref{sec:first_user_study}. In Section \ref{sec:user_study} we present the results and analysis of our third and final user-study for evaluating the technique for redirected walking in VR. 

\section{Related Work}
\label{sec:related_work}
Over the recent years many techniques have been proposed relating to interaction in VR and in particular, navigation. Below we present a brief overview of the state-of-the-art most relevant to our proposed technique. The related work is categorized in terms of (a) redirection and VR, (b) steering algorithms, resetting phase and natural visual suppressions, and (c) dynamic foveated rendering.

\subsection{Redirection and VR}
One of the most important forms of interaction in VR is locomotion. Natural walking is the most preferred (and natural) technique primarily because it provides an increased sense of presence in the VE \cite{usoh1999flying} and improved spatial understanding \cite{peck2011walkInPlace, ruddle2011cognitiveMapping, ruddle2009cognitiveMapping} while reducing the signs of VR sickness \cite{laviola2000cybersickness}. However, the main difficulty of using natural walking as a locomotion technique in VR is the requirement that the size of the physical tracked space (PTS) is comparable in size with the VE, which is often not the case; especially for simulations involving large-scale environments \cite{poullis2009automatic}. To this day it remains a very active area of research with a particular focus on locomotion techniques which do not carry, in any degree, the spatial constraints imposed by the physical space over to the theoretically boundless virtual space.

%Researchers have been trying to tackle this problem since the introduction of trackable VR HMDs. An ample amount of research is being conducted every year to create better locomotion solutions which could overcome this flaw.

In 2001, Razzaque et al. \cite{razzaques2001redirected} proposed the concept of 'Redirected Walking' for the first time. By making subtle rotations to the VE the users were tricked into walking on a curved path in the PTS while maintaining a straight path in the VE. These subtle rotations applied to the VE were enough to convince the users that they had explored a comparatively larger virtual area than the actual available play space. 

Since the original concept of redirected walking, a number of attempts were made to achieve the same effect based on software and/or hardware. Some researchers even tried to achieve this by incorporating physical props \cite{cheng2015turkdeck} or manipulating the entire physical environment \cite{suma2011leveraging,suma2012impossible}. However, these type of solutions failed to gain the mainstream attention due to their many dependencies on factors other than the actions of the user himself. 

Hardware-based approaches were also explored to resolve this problem such as the omnidirectional treadmill \cite{souman2010making}, suspended walking \cite{benjamin2013suspendedWalking}, low-friction surfaces \cite{iwata1996perambulator}, walking in a giant Hamster Ball \cite{medina2008virtusphere}, etc. Souman et al \cite{souman2010making} proposed a technique using an omnidirectional treadmill algorithm to improve the perception of life-like walking in the VE. A position-based controller was used to determine the speed of the user's movements in the VE and to rotate the treadmill accordingly. 
% This speed was determined based on the user's position from a certain reference point on the treadmill. 
Several other attempts were also made to create a perfect omnidirectional treadmill \cite{iwata1999infiniteFloor, darken1997omnitreadmill, nagamori2005ballArrayTreadmill, huang2003strollBased} but they are not considered cost-effective solutions primarily due to the heavy and expensive equipment involved. Following a similar hardware-based approach, a giant hamster-ball-like structure was proposed by Fernandes et al in \cite{fernandes2003cybersphere} where the user was placed in a human-sized hollow sphere which could spin in any direction. Another similar prototype was proposed by Medina et al in \cite{medina2008virtusphere}. These prototypes, although they are possible solutions to the problem of infinite walking in VR, they all lack inertial force feedback. For this reason natural walking is considered to be the most preferred and natural way for locomotion in VR \cite{christensen2000inertial}. Moreover, the multimodal nature of natural walking allows free user movements such as jumping or crouching.

Software-based techniques have also been proposed for achieving the same effect by solely manipulating the VE. These can be divided into two groups: (a) techniques that use the user's head rotations and translations for scaling the VE dynamically based on the scenario as in \cite{mahdi2017toolkit, razzaques2001redirected, razzaques2002redirected}, and  (b) techniques that partially or fully warp the virtual environment \cite{dong2017smooth, sun2016mapping}. 
% More recently, the work of \cite{dong2017smooth, sun2016mapping} has contributed significantly towards improved mapping and rendering techniques since the work of \cite{razzaque2005redirected}. 
Due to the dominance of the visual sense over the other human senses, these techniques focus mainly on reducing the effect of subtle visual distractions resulting from repeated redirection. These visual distractions are mainly created during the naturally occurring or stimulated visual suppressions such as a blink or a saccade. Langbehn et al in one of their recent studies \cite{langbehn2018blinks} proposed a redirection technique which leverages the naturally occurring blinks to redirect the user. The authors performed extensive experiments to determine the thresholds of rotational and translational gains that can be introduced during the blink. Concurrently, Sun et al in \cite{sun2018towards} leveraged the perceptual blindness occurring before, during, and after the saccades to update the environment quickly over several frames. A common disadvantage of these techniques is the fact that they are disruptive to the cognitive task at hand, since they rely on stimulating saccades by introducing artifacts in the rendered image to distract the user's attention. 

\subsection{Steering Algorithms, Resetting Phase and Natural Visual Suppressions}

\textit{Steering Algorithms:} In order to calculate the amount of redirection, two parameters are required: the target direction of the user (a) in the VE, and (b) in the PTS. There are many methods for predicting the target direction in VE ranging from using the user's past walking direction \cite{zank2015pastWalkingDirection}, to head rotations \cite{steinicke2008headRotations}, and gaze direction \cite{zank2015eyeTracking}. As there are spatial constraints in terms of the available PTS, the user must be steered away from the boundaries of the PTS i.e. typically walls. To locate the user's target direction in PTS, Razzaque proposed a number of related agorithms in \cite{razzaque2005redirected}, namely steer-to-center, steer-to-orbit, and steer-to-multiple-targets: (i) steer-to-center steers the users towards the center of the PTS; (ii) steer-to-orbit steers them towards an orbit around the center of the PTS; and (iii) steer-to-multiple-targets steers the users towards several assigned waypoints in the PTS. In \cite{hodgson2013SteerToMultipleCenters}, Hodgson et al proposed the steer-to-multiple-centers algorithm which is essentially an extension of steer-to-center algorithm. In this case, the steer-to-center algorithm proved to be working better than this extension. Another experiment performed by Hodgson et al in \cite{hodgson2014constrainedVE} showed that the steer-to-orbit algorithm worked better in the setting where the directions of walking were limited by virtual objects in the VE. In all of the above cases, the user's direction of movement is constantly changing which implies that the amount of redirection must also be constantly computed. 
% Thus, in order to keep redirecting the user towards a certain point or an orbit, one needs to constantly keep computing the new amount of redirection. 

Recently, Langbehn et al \cite{langbehn2017predefinedCurvedPaths} proposed a redirected walking technique using predefined curved paths in the VE. In this experiment, users were instructed to follow the predefined curves inside the VE and were allowed to change their direction of walking only when they reached the intersection of their current curve with another predefined curve. Since the curves were mapped within the PTS, this eliminated the possibility of crossing over a boundary as long as the user followed the predefined path. 
% As long as the user followed the predefined path, this technique eliminated the chance of bumping into walls as those curves were mapped within the PTS. 
In our work, we employ the \textit{steer-to-center} algorithm for redirecting the user towards the center of the PTS when a collision is predicted to occur.

\textit{Resetting Phase:} The resetting phase is one of the most important aspects of all redirected walking techniques because there is always a small possibility of the user crossing over the boundary of the PTS. If this occurs, the user has to be stopped and their position has to be reset before starting to walk again.

% The technique can be as good as it can be but unless the PTS is equal to the size of VE, there will always be a minor possibility of user hitting the boundary of PTS at some point during the experiment. 

% In this case, the user has to be stopped and reset in a proper way before he can start walking again. 

Several solutions were proposed to address this problem with the most common being in \cite{williams2007resetTechniques}: (i) \textit{Freeze-Turn} is the method where the field-of-view (FoV) of the user remains unchanged while she turns towards the empty space in the PTS, (ii) \textit{Freeze-Back-up} is the method where the FoV remains unchanged while the user backs-up from the boundary making enough room for walking, and (iii) \textit{2:1 Turn} is the method where twice the rotational gain is applied to the user's head rotations, i.e. if the user turns by $180^{\circ}$, a rotation of $360^{\circ}$ is applied so that the user is again facing the same direction that she was facing before. Visual effects which may result from the resetting can be addressed with the use of visual distractors as proposed by Peck et al in \cite{peck2011walkInPlace}.

% In order to minimize the visual effect of the resetting \cite{peck2011walkInPlace} 
% To complement the resetting and minimize The use of visual distractors was made by Peck et al in \cite{peck2011walkInPlace} to minimize the visual effect of the reset. 

Furthermore, most of the redirected walking techniques (e.g. Redirected Walking Toolkit \cite{mahdi2017toolkit}) follow the stop-and-go paradigm where if the user crosses over a boundary, and before she starts walking again, she will have to perform an in-situ rotation towards a direction where there is no obstacle. We follow the same paradigm in our work.
% We also use the same approach to reset the user in our system.

\textit{Natural Visual Suppressions:} The human visual system is not perfect. Due to several very frequent involuntary actions, humans face temporary blindness for short periods of time \cite{regan2000temporaryBlindness, rensink1997temporaryBlindness, rensink2002temporaryBlindness}. These involuntary actions are called visual suppressions. Saccades, eye-blinks, the phase of nystagmus, and vergence movements are some of the involuntary visual suppressions \cite{volkmann1986suppressions}. Saccades are the brief rapid eye movements that occur when we quickly glance from one object to another; eye-blink is a rapid opening and closing of the eyelids - these eye movement can occur either voluntarily, involuntarily or as a reflex; the phase of nystagmus is a condition where uncontrolled rapid eye movements occur from side-to-side, top-to-bottom or in circular motion; vergence movement occurs to focus on the objects with different depths, one after the other \cite{volkmann1986suppressions}. In the following, we review only techniques employing saccades and eye-blinks since our proposed technique only utilizes these.

% \footnote{WE ARE USING BOTH BLINKS AND SACCADES TO UPDATE THE USER'S FOVEAL ZONE.}. 
Saccades are extremely unpredictable, rapid, and ballistic eye movements that occur when we abruptly shift our fixation point from one object to the other \cite{bahill1975saccadeSpeedRange}. The visual suppression occurs before, during, and after the saccadic movement \cite{burr1994saccadeTimings} and could last for 20-200ms while the speed of these saccadic movements can reach up to 900$^{\circ}$/s \cite{bahill1975saccadeSpeedRange}. Researchers are still trying to identify an exact reasoning behind this involuntary human behavior \cite{burr1994saccadeTimings, ibbotson2009unknownReasoning}. A saccade occurs very frequently and can last for several frames \cite{langbehn2018blinks} which makes it possible to render the updated VE without the user noticing. 

In contrast to the saccades, blinks occur much less frequently and are much slower which gives more flexibility for reorientation due to the longer induced change blindness \cite{regan2000temporaryBlindness}. Depending upon the scenario, one blink can give the users a temporary blindness of 100-400 ms which is much longer than a saccade \cite{ramot2008blinkDuration}. This makes them easier to detect even with readily available off-the-shelf eye trackers. Similar to saccades, our visual perception is also suppressed before, during, and after the opening and closing movements of the eyelids \cite{volkmann1980blinks, volkmann1986suppressions}. A study performed by Bentivoglio in \cite{bentivoglio1997blinkrate} showed that an average person blinks at an average rate of 17 times per minute. 

In our work, we leverage this physiological phenomenon to refresh the foveal zone render and therefore redirect the user multiple times per minute during blinks. Furthermore, we leverage information reported in recent studies for determining the maximum rotational and translational thresholds for VR during blinks and saccades \cite{langbehn2018blinks, sun2018towards, langbehn2016blinksThreshold, benjamin2015saccadeThresholds} to update the VE and refresh the render without the user perceiving anything.

\subsection{Dynamic Foveated Rendering}

% With VR technologies getting better and better each year, we need improved processing power to render all those high quality frames at a rate of at least 90 frames per second (fps) to provide the users with a smooth experience. The problem in this is with increasing device performance, the hardware becomes expensive and not everyone can afford it. 

The concept of foveated rendering was first introduced by Reder in 1973 \cite{reder1973FirstFoveatedRender}. This is a technique which drastically reduces the overall GPU workload while providing the same VR experiences as before. Foveated rendering leverages the fact that small changes occurring in our peripheral vision are imperceptible to humans. Thus, the area of the image corresponding to the peripheral vision can be rendered at a much lower resolution while the area of the image corresponding to the foveal vision is rendered at full resolution \cite{patney2016perceptuallyBasedFR}.

% Foveated rendering leverages this information, and more recently user's eye movements, to render only the parts in screen space which are most important or where the gaze of the user is fixated with full resolution while reducing the resolution of rest of the display significantly \cite{patney2016perceptuallyBasedFR}. 

% Since Reder, there have been several attempts at foveated rendering \cite{guenter2012foveatedRenderingOld, duchowski2007foveatedRenderingOld, zha1999foveatedRenderingOld}. An attempt at perceptually-guided foveated rendering was also made by Luebke and Hallen in \cite{luebke2001peceptuallyGuidedFR}. 

Although in recent years researchers have proposed many software-based techniques for simulating perceptually-guided foveated rendering e.g. \cite{yongHe2014multiRateShading, stengel2016peceptuallyGuidedFR, sun2018towards, patney2016perceptuallyBasedFR}, it was until the very recent introduction of Nvidia's Variable Rate Shading \cite{nvidiaVRS} that foveated rendering was supported in hardware and integrated into the rendering pipeline. This integration has reduced latency and aliasing effects close to zero as it was demonstrated by the real-time visualization platform called ZeroLight \cite{zerolightVRS}. 

% followed a more practical approach towards foveated rendering in modern VR headsets. 

% And in the more recent work, researchers focused on implementing this technique on modern HMDs with inbuilt gaze trackers \cite{tursun2019foveatedRendering, patney2016gazeTrackedFR}. Just like Luebke and Hallen, attempts were also made towards perceptually-guided foveated rendering for modern VR headsets \cite{stengel2016peceptuallyGuidedFR, sun2018towards, patney2016perceptuallyBasedFR}.

% \begin{figure}[!ht]
%     % \captionsetup{justification=centering}
%     \centering
%     \includegraphics[width=0.47\textwidth]{images/vrs_architecture.png}
%     \caption{NVIDIA's Variable Rate Shading architecture}
%     \label{fig:vrs_architechture}
% \end{figure}

%  Figure \ref{fig:vrs_architechture} shows the architecture of VRS \cite{nvidiaVRSarchitecture}. 
 
% Recently, NVIDIA introduced Variable Rate Shading (VRS) which shades individual pixels based on their degree of importance \cite{nvidiaVRS}. In addition to this, as the tracking system of VR headsets detects and interpolates even the smallest head movements constantly, it is prone to noticeable aliasing effects. VRS can also be used to apply higher resolution than whats maximum supported for the display in the foveal zone. This type of rendering technique is known as super-sampling and it can drop the aliasing effect close to zero. A recent practical example of this technique is demonstrated by ZeroLight in their real-time visualization platform \cite{zerolightVRS}. 

In our work we employ Nvidia's VRS not only for reducing the overall GPU workload but also for blending foveal and non-foveal (peripheral) zones rendered from two co-located cameras, respectively.

% to reduce the overall GPU workload and ensure a real-time performance, and more importantly for rendering foveal and peripheral

\begin{figure*}[!ht]
    % \captionsetup{justification=centering}
    \centering
    \includegraphics[width=0.8\textwidth]{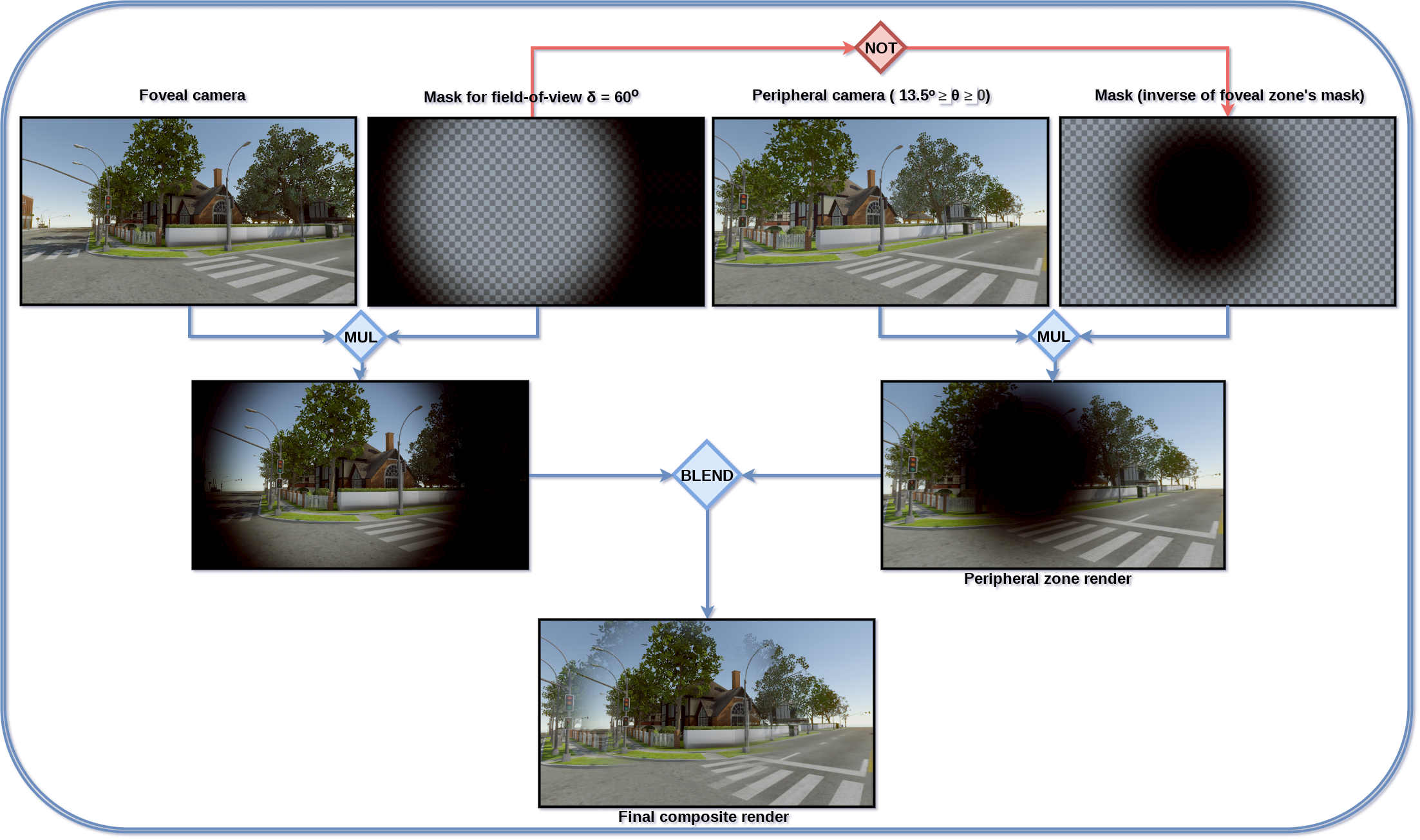}
    \caption{The pipeline of the proposed technique involves rendering the VE from two co-located cameras $Cam_{foveal}$ with field-of-view $\delta=60^{\circ}$ and $Cam_{non-foveal}$ with rotation angle $13.5^{\circ} > \theta > 0^{\circ}$. As demonstrated by the results of user study \#2 the users fail to perceive any visual distractions or artifacts in the final composite render while they are preoccupied with a cognitive task; which is almost always the case with VR applications.}
    \label{fig:system_overview}
\end{figure*}

\section{Technical Overview}
\label{sec:system_overview}

Figure \ref{fig:system_overview} shows the pipeline of the proposed technique. Two co-located cameras $Cam_{foveal}, Cam_{non-foveal}$ render the VE. Based on the results of the first user study we have determined that the field-of-view for $Cam_{foveal}$ is $\delta = 60^{\circ}$, and the rotation angle applied to the VE and rendered from $Cam_{non-foveal}$ is $13.5^{\circ} > \theta > 0^{\circ}$. A mask corresponding to $\delta = 60^{\circ}$ is applied on the rendered image of $Cam_{foveal}$, and the inverse of the same mask is applied on the rendered image of $Cam_{non-foveal}$. The resulting masked renders are then composited into the final render displayed to the user. As demonstrated by the results of the user studies \#2 and \#3, the users fail to perceive any visual distractions or artifacts in the final composite render while they are preoccupied with a cognitive task; which is almost always the case in VR applications.

\section{User Study \#1: Determining Maximum Rotation Angle and Field-of-View of Foveal Zone}
\label{sec:first_user_study}

The efficacy of redirected walking is tightly coupled with the user's perception of the redirection taking place. In our first user-study we determine the maximum angle for which the rotation of the non-foveal zone (i.e. the area in the rendered image corresponding to the peripheral vision) remains imperceptible by the user. 

%The effectiveness of foveated redirection depends on several factors such as redirection frequency, size of the foveal region, and visual perception of the offset angle between this region and the peripheral background. These factors had to be predetermined as they portray a major significance during the redirection. 

\subsection{Application}
\label{subsec:first_user_study_application}

\begin{figure}[!ht]
    % \captionsetup{justification=centering}
    \centering
    \includegraphics[width=0.5\textwidth]{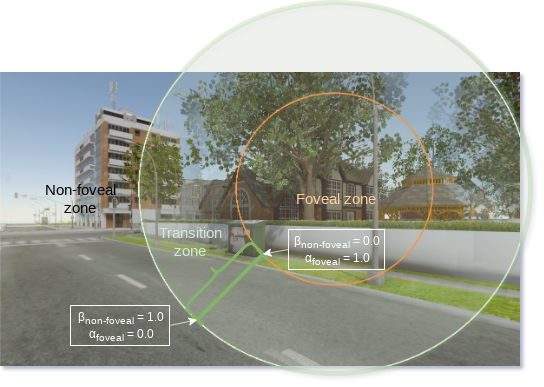}
    \caption{User's perspective during the user study \#1. The foveal zone marked with the orange circle is rendered at the high quality (1:1 sampling). The non-foveal zone is the complement of the foveal zone and is rendered at a lower resolution (16:1 sampling)}
    \label{fig:pilotstudy_screensnap}
\end{figure}

We designed an immersive VR application using the HTC Vive Pro Eye HMD with an integrated Tobii Eye Tracker. The application depicted a serene urban city environment in which red spheres were hidden at random locations. The environment was developed in Unity3D and foveated rendering was supported using an NVIDIA RTX 2080Ti graphics card. Three zones were specified for the foveated rendering:

% each with different subsampling schemes: (1) the foveal zone where pixels were rendered with a 1:1 sampling, (2) the non-foveal (or peripheral) zone where pixels were rendered with a 16:1 sampling, and (3) the overlapping zone where pixels were rendered with a 4:1 sampling. The diagram in Figure \ref{subsampling_scheme_for_zones} shows the subsampling schemes used for each of the zones. 

\begin{enumerate}
\item \textbf{Foveal zone:} The foveal zone is the circular area in the rendered image centered at the fixation point captured by the eye tracker.  For rendering, this zone must have the highest quality since this is where the user's foveal vision is focused. Thus, pixels in the foveal zone are rendered with a 1:1 sampling.

\item \textbf{Non-foveal zone:} The non-foveal zone is the area in the rendered image (complementary to the foveal zone) which corresponds to the peripheral vision. This zone is of lower importance than the foveal zone since it is not the main focus of the user's vision. Hence, pixels in the non-foveal zone are rendered with a 16:1 sampling.

\item \textbf{Transition zone:} The transition zone is the overlapping area of the foveal and non-foveal zones. This zone was introduced after initial experiments had shown that having only the foveal and non-foveal zones results in sharp boundary edges on the circular area separating them which are immediately perceived by the user. Pixels in the transition zone are rendered with a 4:1 sampling.

\end{enumerate}

\noindent
Figure \ref{fig:pilotstudy_screensnap} shows a frame \footnote{The visible frame in VR is panoramic and considerably larger than the one shown here. We are showing only the part of the frame relevant to the discussion.} from the application with the three zones annotated. The foveal zone corresponding to a field-of-view of $\delta_{foveal} = 60^{\circ}$ is marked with an orange circle and is rendered at the highest quality with 1:1 sampling and no rotation. The non-foveal zone is rendered at a lower resolution with 16:1 sampling and a rotation of $\theta_{non-foveal} = 6^{\circ}$. 

The transition zone is also shown as the green ring around the foveal zone, rendered with 4:1 sampling. This indicates the overlapping area between the foveal and non-foveal zones for which alpha-blending is performed at each pixel with $C_{blended} = \alpha_{foveal} \times C_{foveal} + \beta_{non-foveal} \times C_{non-foveal}$ where $\beta_{non-foveal} = 1.0 - \alpha_{foveal}$, $C_{blended}$ is the resulting blended color, and $C_{foveal}, C_{non-foveal}$ are the foveal and non-foveal colors at a pixel, respectively. This requires two co-located cameras (with different rotations) in order to render the foveal and non-foveal frames from which the color values $C_{foveal}$ and $C_{non-foveal}$ are retrieved. The boundary values $[0.0, 1.0]$ for $\alpha_{foveal}$ and $\beta_{non-foveal}$ are also shown in this figure. These coincide with the boundaries of the transition zone. The field-of-view $\delta_{transition}$ corresponding to the transition zone is defined empirically as an offset to the field-of-view $\delta_{foveal}$ of the foveal zone, given by $\delta_{transition} = \delta_{foveal} + 40^{\circ}$.

\subsection{Procedure}
\label{subsec:first_user_study_procedure}
The premise of our hypothesis is inattentional blindness which implies that the user's attention must be directed towards another cognitive task. Thus, we instructed the participants to perform a target-retrieval task. More specifically, the participants were asked to search and count red spheres hidden in the VE. At each iteration of the experiment, the red spheres were hidden at random locations. This was done in order to eliminate the possible bias that may be introduced by memorizing the locations between iterations.

The first user study involved 11 participants (2 females, 18.2\%). Average age was 24.64 with a SD of 2.8. Median of their reported experiences with using VR devices was 3, and the median of their experiences with using an eye tracking device was also 3 on a 5-point Likert scale, with 1 being least familiar, and 5 being most familiar. The participants performed this task from a seated position and were only allowed to rotate their chair in-place. The participants were instructed to press the trigger button on the Vive controller if and when they noticed a visual distortion or felt nausea due to simulator sickness. During the experiment, the rotation angle $\theta$ of the non-foveal zone was gradually increased and grouped by increasing order of the field-of-view $\delta$ of the foveal zone i.e. ($[(\delta_{1}, \theta_{1}), (\delta_{1}, \theta_{2}), \dots, (\delta_{1}, \theta_{n}), (\delta_{2}, \theta_{1}), \dots, (\delta_{2}, \theta_{n}), \dots]$). Each time the trigger was pressed, the pair of $(\delta_{i}, \theta_{i})$ was recorded, and then the experiment continued with the now increased field-of-view $\delta_{i+1}$ and a reinitialized rotation angle $\theta_{1}$. 

The range of values for the field-of-view was $20^{\circ}$ to $60^{\circ}$.
% ; a typical range for VR applications. 
The step of each increment was $10^{\circ}$ after the completion of one cycle of the rotation angle, or until triggered by the user. During a cycle, the rotation angle ranged from $0^{\circ}$ to $15^{\circ}$ and the step of each increment was $1^{\circ}$ per second. 

Preliminary experiments during the design of the application had shown that repeated increments of the field-of-view of the foveal zone can lead to nausea and severe dizziness. For this reason, the participants were instructed to take a short break after each cycle of increments of the field-of-view. Furthermore, the sequence of the cycles i.e. the field-of-view values, was randomized for each participant in order to eliminate any bias. Figure \ref{fig:pilotstudy_screensnap} shows the view from the user's perspective during the experiment.

\subsection{Analysis of results}
\begin{figure}[!ht]
    % \captionsetup{justification=centering}
    \centering
    \includegraphics[width=0.47\textwidth]{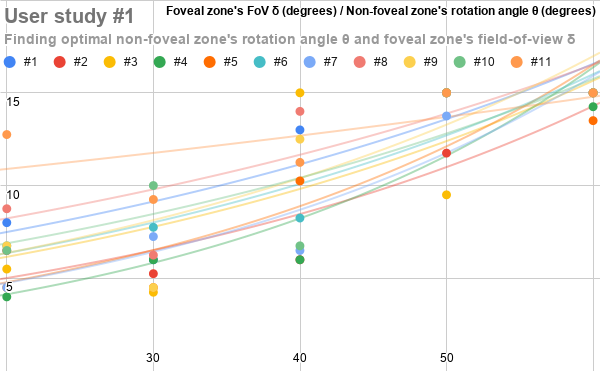}
    \caption{Participants' responses for pairs of $(\delta, \theta)$. We select the \textit{smallest rotation} angle for which users did not perceive a change associated with the \textit{largest} field-of-view for which the majority of the users did not perceive a change (i.e. 9 out of 11). Thus, the optimal pair values for $(\delta, \theta)$ is determined to be $(60^{\circ}, 13.5^{\circ})$. The exponential trendlines are also shown which confirm that as the field-of-view $\delta$ increases, the tolerance for higher rotation angle $\theta$ also increases.}
    \label{fig:user_study1}
\end{figure}

The results are shown in Figure \ref{fig:user_study1}. A cycle of the rotation angle $\theta \in [0^{\circ}, 15^{\circ}]$ was performed for each field-of-view $\delta \in [20^{\circ}, 60^{\circ}]$. The results show that as the field-of-view $\delta$ increases the tolerance for higher rotation angle $\theta$ also increases, which can also be confirmed by the exponential trendlines shown for each participant. For the reasons mentioned above, we select the \textit{smallest rotation} angle for which users did not perceive a change associated with the \textit{largest} field-of-view for which the majority of the users did not perceive a change (i.e. 9 out of 11). Thus, the ideal pair values for $(\delta, \theta)$ is determined to be $(60^{\circ}, 13.5^{\circ})$; where $13.5^{\circ}$ is the maximum allowed rotation angle.

\subsubsection{Simulator Sickness Questionnaire (SSQ)}
\begin{figure}[!ht]
    \centering
    \includegraphics[width=0.47\textwidth]{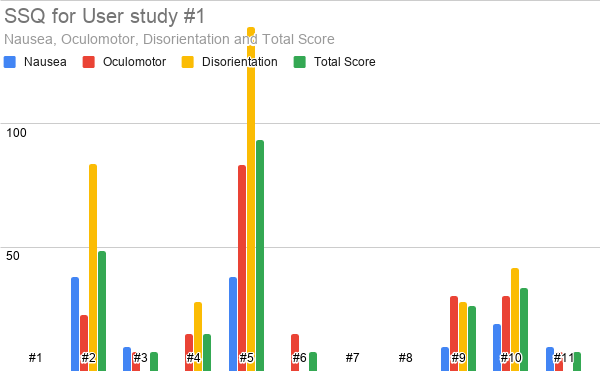}
    \caption{Results from the SSQ scores (Left to right: Nausea, Oculomotor, Disorientation and Total Severity). The Total Severity and sub-scales such as Nausea, Oculomotor, and Disorientation were calculated based on the formulas in \cite{kennedy1993ssq}}
    \label{fig:SSQAnalysis}
\end{figure}

Upon completing the experiment all participants were asked to complete Kennedy Lane's Simulation Sickness Questionnaire (SSQ). The Total Severity (TS) and the sub-scales Nausea, Oculomotor, and Disorientation were calculated using the formulas from \cite{kennedy1993ssq}. Based on the SSQ categorization provided by Kennedy et al. in \cite{kennedy2003ssq}, $55\%$ of the participants reported no signs (TS=0) or minimal signs (TS$<10$) of simulator sickness. All the participants completed the test, with $0$ dropouts. Upon further analysis, the disorientation sub-scale had the highest average score of $29.11$ with a maximum score of $139.2$. This was expected, considering the fact that the rotation angle was constantly increasing and thus the VE rendered in the HMD induces conflicts between the vestibular and visual signals, leading to vestibular disturbances such as vertigo or dizziness. The results from SSQ responses are summarized in Table \ref{tab:SSQ} and Figure \ref{fig:SSQAnalysis}.

% leading to vestibular disturbances since the observed as the HMD induces conflicts between our vestibular and visual signals

% until the users perceive a change which causes the disorientation. The disorientation sub-scale includes symptoms such as vertigo or dizziness which are caused by vestibular disturbances. These vestibular disturbances are observed as the HMD induces conflicts between our vestibular and visual signals. The results from SSQ responses are shown in Table \ref{tab:SSQ} and Figure \ref{fig:SSQAnalysis}.

\begin{table}[!ht]
  \begin{center}
    \caption{Results from the responses of SSQ. The Total Severity (TS) and the corresponding sub-scales such as Nausea, Oculomotor, and Disorientation were calculated using the formulas from \cite{kennedy1993ssq}. The majority (55\%) of the participants reported no signs (TS=0) or minimal signs (TS$<10$) of simulator sickness. Highest average score for disorientation as expected.}
    \label{tab:SSQ}
    \begin{tabular}{|l|c|c|c|c|c|}
      \hline
      \textbf{Scores} & \textbf{Mean} & \textbf{Median} & \textbf{SD} & \textbf{Min} & \textbf{Max}\\ % <-- added & and content for each column
      \hline
      \textbf{Nausea (N)} & 11.27 & 9.54 & 14.67 & 0 & 38.16\\ % <--
      \hline
      \textbf{Oculomotor (O)} & 19.29 & 15.16 & 24.06 & 0 & 83.38\\ % <--
      \hline
      \textbf{Disorientation (D)} & \textbf{29.11} & 0 & 45.09 & 0 & \textbf{139.2}\\ % <--
      \hline
      \textbf{Total Score (TS)} & 21.76 & 7.48 & 28.47 & 0 & 93.5\\ % <--
    %   \textbf{N} & 11.27 & 9.54 & 14.67 & 0 & 38.16\\ % <--
    %   \textbf{O} & 19.29 & 15.16 & 24.06 & 0 & 83.38\\ % <--
    %   \textbf{D} & 29.11 & 0 & 45.09 & 0 & 139.2\\ % <--
    %   \textbf{TS} & 21.76 & 7.48 & 28.47 & 0 & 93.5\\ % <--
     \hline
    \end{tabular}
  \end{center}
\end{table}

\subsection{Discussion}
The design of the application of user study \#1 involved making a decision on the range of values for (a) the field-of-view $\delta$ for the foveal zone, and (b) the rotation angle $\theta$ of the non-foveal zone:
\begin{itemize}
\item \textbf{Range of $\delta$:} Humans have a maximum horizontal field-of-view of about $210^{\circ}$. This is further reduced to $110^{\circ}$ by the hardware i.e. maximum field-of-view for HTC Vive Pro. 

The foveal and non-foveal zones are by definition complementary to each other. Thus, there is a tradeoff between the sizes of the foveal and non-foveal zones. If $\delta$ is large, then the foveal zone is large, and the non-foveal zone is small. Having a small non-foveal zone means that only a small area of the final composite image will be rendered from the non-foveal camera $Cam_{non-foveal}$, as shown in Figure \ref{fig:system_overview}, leading to smaller possible redirections. When $\delta = 60^{\circ}$ the foveal zone occupies $54.55\%$ of the final composite render, and the non-foveal zone occupies $45.45\%$ (including a transition zone of $35.45\%$). Similarly, if $\delta = 90^{\circ}$ the foveal zone occupies $90.91\%$ of the final composite render which does not leave much for the non-foveal zone. In contrast, if $\delta$ is small, then the foveal zone is small and the non-foveal zone is large. Although this allows for larger redirections, we have found in our preliminary tests that when $\delta < 20^{\circ}$ it can cause severe nausea and simulation sickness. 

For these reasons we have selected the range of values $\delta \in [20^{\circ}, 60^{\circ}]$ which balances the sizes of the foveal and non-foveal zones, and is large enough that it does not cause user discomfort. 

\item \textbf{Range of $\theta$:} Recent experiments reported in \cite{sun2018towards} have shown that users cannot tolerate a rotational angle of more than $12.6^{\circ}$ in their field-of-view during a saccade having a velocity of $180^{\circ}/sec$. Based on this, we have selected the range $\theta \in [0^{\circ}, 15^{\circ}]$.

\end{itemize}

\section{User Study  \#2: In-situ Gaze Redirection using Dynamic Foveated Rendering}
% A second user study was conducted to confirm the results of the first user study on the optimal pair values being $(\delta = 60^{\circ}, \theta=13.5^{\circ})$. More specifically, 

The objective of the second user study is twofold:
\begin{enumerate}
    \item Firstly, to determine whether a rotation of the VE by an angle below the maximum (i.e. $\theta < 13.5^{\circ}$) is indeed imperceptible and does not cause simulation sickness.
    \item Secondly, to experimentally prove with quantitative measures that using the proposed redirection technique with gaze-only (and without walking) the rotation of the VE results in the equivalent redirection of the participant in the PTS.
\end{enumerate}
 
\subsection{Application and Procedure} 
An experiment was devised similar to the one in Section \ref{sec:first_user_study}. In contrast to the user study \#1, the participants were only allowed to spin in-situ from a \textit{standing position}. This change from a seated to a standing position eliminates the possibility of the participants perceiving any orientation and navigation cues coming from the orientation of the chair. The participants were given a target-retrieval task and instructed to retrieve, by directing their gaze to, as many virtual targets (i.e. orange pumpkins) as possible. The virtual targets disappeared as soon as the gaze direction intersected their bounding box. The positions of the targets were randomized for each participant.

The duration of the experiment was 60 seconds. Unbeknownst to the participants, in the first 30 seconds there was no redirection applied i.e. $\theta_{Cam_{non-foveal}} = 0$. This served as a baseline for participants who had little to no experience in using HMDs. Once accustomed to the VE, during the following 30 seconds the rotation angle of the VE was increased at a rate of $6^{\circ}/s$. 
% which is below the maximum rotation angle of $13.5^{\circ}$ determined in user study \#1. 

Hence, the hypothesis is that after 30 seconds of a constant smooth rotation of the VE at a moderate rate of $6^{\circ}/s$ the participant should face $180^{\circ}$ away from their initial orientation i.e. the opposite direction. To prove this, the initial (i.e. at $time=0s$) and the final (i.e. at $time=60s$) gaze directions of each participant were recorded. Additionally, before each participant removed the HMD at the end of the experiment they were asked to face towards what they believed to be their initial directions in the PTS using visual landmarks from within the VE to orient themselves.

\subsection{Analysis of Results}
The study involved 11 participants (2 females: 18.2\%, average age of $26.27 \pm  3.13$). Based on a 5-point Likert scale the medians of their experience with using VR or any other eye tracking devices were 3. Five of the participants had not taken part in user study \#1 (Participants \#2, \#3, \#6, \#9, \#11). After the experiment, the participants completed the SSQ.
% \cite{kennedy1993ssq}.
% which reported an improved overall user experience. 
The angle between the initial and final gaze directions was calculated for each participant. The average deviation 
% The average deviation between the initial and fina gaze directions was On an average, every participant's final gaze direction 
was $171.26^{\circ}$ (4.77 SD) which means that the participants thought that their initial orientation was towards the opposite direction. In fact, all participants reported that they did not perceive the redirection and were surprised by how off their 'sensed' orientations were.
% Results from this experiment are shown in Figure \ref{fig:SSQ2Analysis}.

% \begin{figure}[!ht]
%     \centering
%     \includegraphics[width=0.47\textwidth]{images/user_study_2.png}
%     \caption{}
%     \label{fig:user_study2}
% \end{figure}

\subsubsection{Simulator Sickness Questionnaire (SSQ)}

\begin{figure}[!ht]
    \centering
    \includegraphics[width=0.47\textwidth]{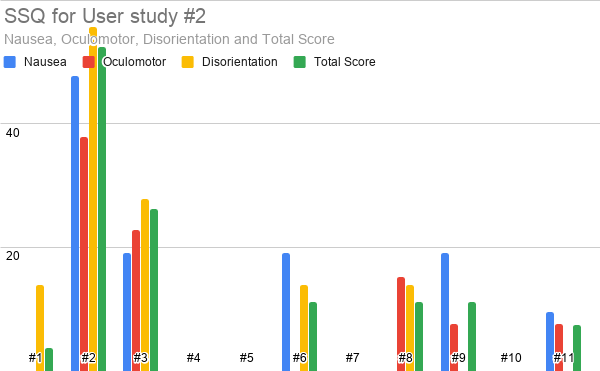}
    \caption{Results from the SSQ scores (Left to right: Nausea, Oculomotor, Disorientation and Total Severity). The Total Severity and sub-scales such as Nausea, Oculomotor, and Disorientation were calculated based on the formulas in \cite{kennedy1993ssq}}
    \label{fig:SSQ2Analysis}
\end{figure}

Based on the scores reported by the participants in the post-test SSQ, the majority of the participants (55\%) showed no signs (TS=0) or minimal signs (TS$<10$) of simulator sickness. The highest score and average was reported for the sub-scale disorientation although reduced by a factor of 2 from user study \#1. This was anticipated since the rotation angle was less than the maximum determined from user study \#1. As it can be seen from Figure \ref{fig:SSQ2Analysis}, one of the participants (\#2) had no previous experience with VR and reported perceptual anomalies including difficulty concentrating, fullness of head and difficulty focusing. 
% had the highest score and highest average score among other sub-scales but this time both were reduced by a factor of more than 2 from the first study. 
% The average Total Severity had also dropped almost by a factor of 2 (from 21.76 to 11.22) which proves that in general, all the users were more comfortable to experience this test as compared to the test in first study.
The results for the SSQ are summarized in Table \ref{tab:SSQ2}.
% and Figure \ref{fig:SSQ2Analysis}. 

% The study was conducted 10 days after the first (in the case of overlap)

\begin{table}[!ht]
  \begin{center}
    \caption{Results from the responses of SSQ for user study \#2.}
    \label{tab:SSQ2}
    \begin{tabular}{|l|c|c|c|c|c|}
      \hline
      \textbf{Scores} & \textbf{Mean} & \textbf{Median} & \textbf{SD} & \textbf{Min} & \textbf{Max}\\ % <-- added & and content for each column
      \hline
      \textbf{Nausea (N)} & 10.41 & 0 & 15.06 & 0 & 47.7\\ % <--
      \hline
      \textbf{Oculomotor (O)} & 8.27 & 0 & 12.43 & 0 & 37.9\\ % <--
      \hline
      \textbf{Disorientation (D)} & \textbf{11.39} & 0 & 17.41 & 0 & \textbf{55.68}\\ % <--
      \hline
      \textbf{Total Score (TS)} & 11.22 & 7.48 & 15.78 & 0 & 52.36\\ % <--
    %   \textbf{N} & 11.27 & 9.54 & 14.67 & 0 & 38.16\\ % <--
    %   \textbf{O} & 19.29 & 15.16 & 24.06 & 0 & 83.38\\ % <--
    %   \textbf{D} & 29.11 & 0 & 45.09 & 0 & 139.2\\ % <--
    %   \textbf{TS} & 21.76 & 7.48 & 28.47 & 0 & 93.5\\ % <--
     \hline
    \end{tabular}
  \end{center}
\end{table}

\begin{figure*}[!ht]
    \centering
% \captionsetup{justification=centering}
    \begin{subfigure}[t]{0.49\textwidth}
        \centering
        \includegraphics[width=\textwidth]{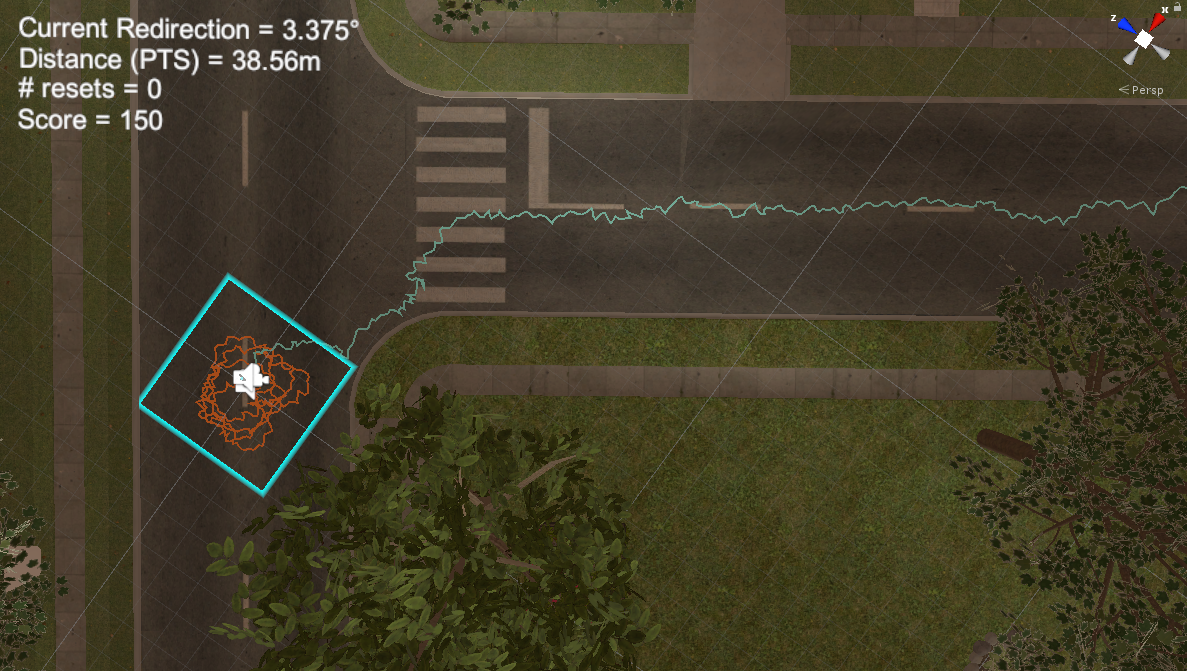}
    \end{subfigure}%
    ~
    \begin{subfigure}[t]{0.49\textwidth}
        \centering
        \includegraphics[width=\textwidth]{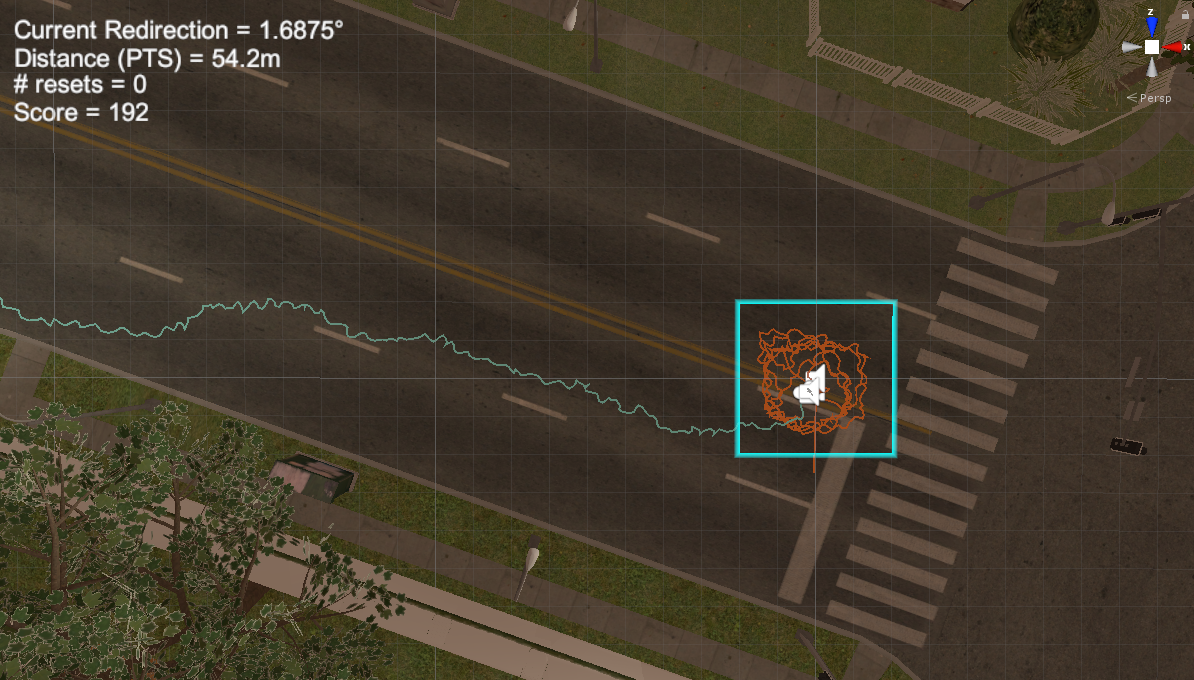}
    \end{subfigure}%
    \caption{Examples from two participants' tests of the experimental condition of user study \#3. The path walked in PTS up to that point is shown in orange color. The corresponding path in the VE is shown in blue color. The cyan-colored box indicates the $4\times4m^2$ available PTS and the camera icon inside the box indicates the location of the user w.r.t. the PTS. Statistics are shown at the top left corner about the current redirection (measured in degrees), distance traveled in PTS (measured in meters), number of resets required, and the score at that point in time. For safety reasons the resetting mechanism of 2:1 was implemented. Steer-to-center algorithm was used for redirecting.}
    \label{fig:paths}
\end{figure*}

\section{User Study \#3: Redirected Walking using Dynamic Foveated Rendering}
\label{sec:user_study}

The objective of the third user study is to evaluate the efficacy of redirected walking during inattentional blindness using dynamic foveated rendering. 

\subsection{Application}
\begin{figure}[!ht]
    \centering
    \includegraphics[width=0.47\textwidth]{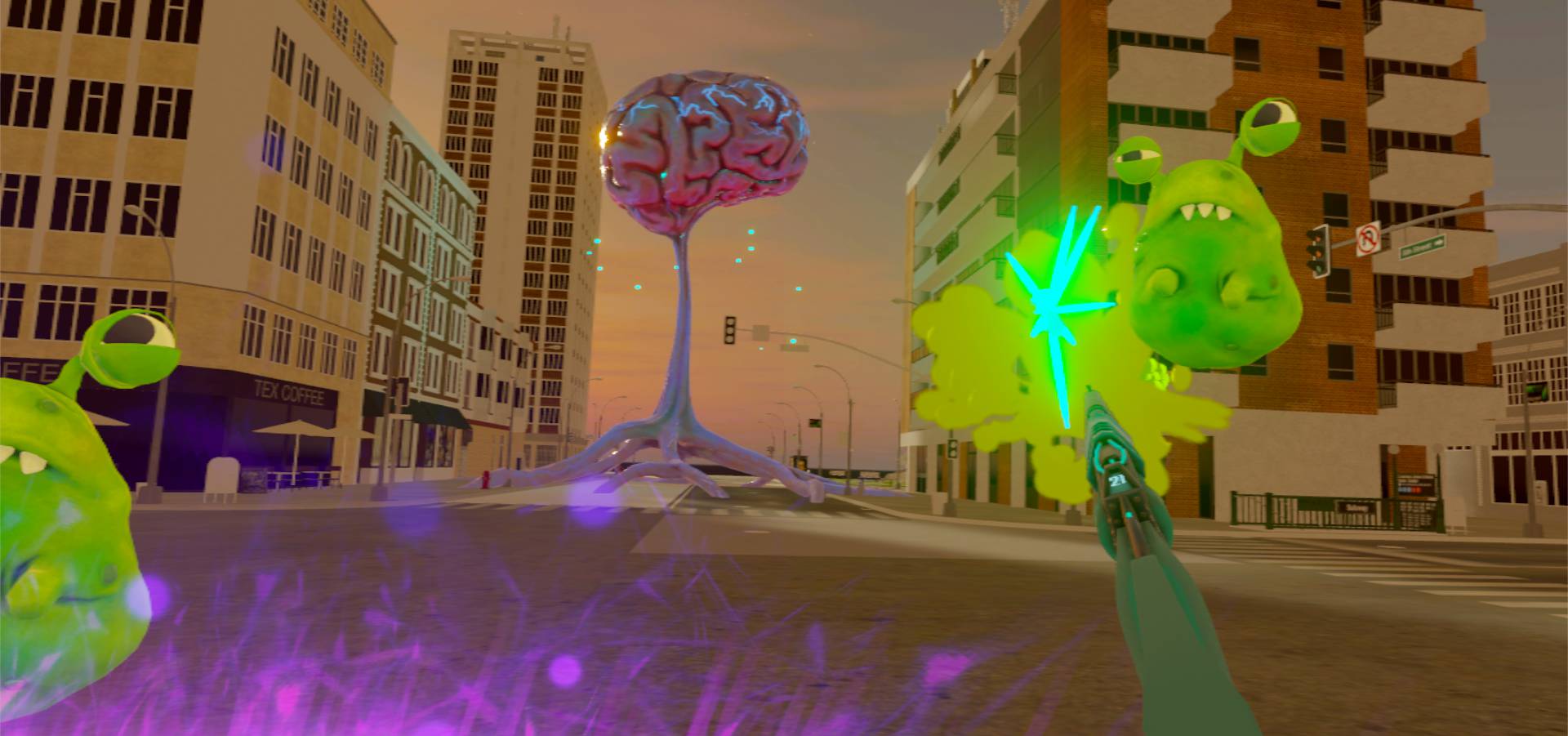}
    \caption{The game designed for user study \#3. The objective is to stop an alien invasion by walking to a predefined location in the VE and destroying the alien-mothership (appearing in the form of a giant brain) while shooting aliens along the way.}
    \label{fig:gameplay_brain_alien}
\end{figure}

As previously explained, inattentional blindness refers to the inability of an individual to see a salient object in plain sight due to lack of attention. This is true for the majority of the VR applications where the user is actively engaged and preoccupied with a cognitive task e.g. games, training simulations, etc. Thus, for the purposes of this user study, we designed a first-person VR game where the objective is to stop an alien invasion. To achieve this the user has to walk through a deserted urban city to a predefined location indicated by a large purple orb and destroy the alien-mothership (appearing in the form of a giant brain) while zapping green aliens along the way. Zapping one alien will award one score point to the player. The green alien enemies are randomly spawned (and are therefore independent of the orientation of the current redirection) only \textit{within} the field-of-view of the user while also making a sound effect. An example of in-game gameplay is shown in Figure \ref{fig:gameplay_brain_alien}. 

The shortest distance the participants had to travel in the VE was $42m$ while the available PTS has a size of $4 \times 4m^2$. The PTS is shown as the cyan-colored box in Figure \ref{fig:paths} and the position of the user w.r.t. the PTS is indicated with the camera icon. For safety reasons, a resetting mechanism of 2:1 was implemented. In cases where the game predicts that the user is about to cross over a boundary of the PTS, it would pause and prompt the user to rotate in-situ by $180^{\circ}$. During the user's rotation, the VE was also rotated by the same angle but in the opposite direction. The user was then allowed to proceed with playing the game.

Redirection was primarily performed by blending in real-time the foveal and non-foveal renders. Furthermore, redirection was also performed during the tracked \textit{naturally occurring} blinks and saccades. In contrast to the state-of-the-art \cite{sun2018towards}, our approach does not stimulate saccades nor blinks since these are disruptive to the cognitive task at hand. 

\subsection{Procedure}
In order to evaluate the efficacy of our technique, a controlled experiment is conducted where the independent variable being tested is the proposed redirection technique. The participants were instructed to complete the objective of the game twice: the first time with the experimental condition i.e. with redirected walking, and after a short break a second time with the control condition i.e. without redirected walking. For the experiment with the control condition, participants had to navigate themselves to the destination by relying solely on the resetting mechanism every time they went out-of-bounds from the available $4\times4m^2$ PTS.

A sample size estimation with an effect size of 0.25 showed that a total of 25 participants were required for the experiment. All the participants were randomly chosen (12\% female, average age of 25.88 years with a SD of 3.06). Based on a 5-point Likert Scale, the median of their experiences using VR headsets or any other eye tracking devices was 3. 

Before the experiment, participants were briefed on their objective. Instructions were also given on how the resetting mechanism works in case they are prompted with an out-of-bounds warning and are required to reset their orientation. Moreover, they were instructed to walk at a normal pace which will allow them to complete the task along the way. 
% Additionally, for safety purposes, they were also instructed to walk slowly during the entire experiment. 
Once both the objectives were completed, participants were also asked to complete the SSQ \cite{kennedy1993ssq}. Furthermore, for the experimental condition, at the end of the first experiment the participants were asked "Did you feel the redirection or any other scene or camera modulation during the experience?".

% The final user study consisted of two experiments. All the participants were asked to perform each experiment once. First experiment included the proposed redirected walking system while the other experiment only relied on player reset to make walking space in front of the player. 

\begin{figure}[!ht]
    \centering
% \captionsetup{justification=centering}
    \begin{subfigure}[t]{0.16\textwidth}
        \centering
\includegraphics[width=\textwidth]{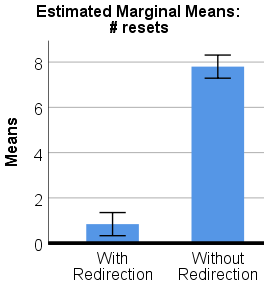}
    \end{subfigure}%
    ~
    \begin{subfigure}[t]{0.16\textwidth}
        \centering
        \includegraphics[width=\textwidth]{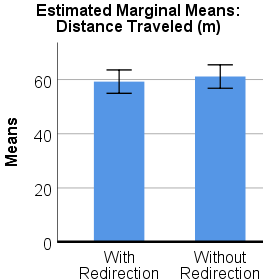}
    \end{subfigure}%
    ~ 
    \begin{subfigure}[t]{0.16\textwidth}
        \centering
        \includegraphics[width=\textwidth]{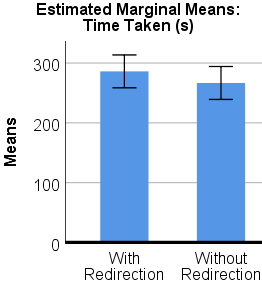}
    \end{subfigure}%
%    \begin{subfigure}[t]{0.24\textwidth}
%        \centering
%        \includegraphics[width=\textwidth]{images/Graphs/ANOVA_Horizontal/Scores_Anova_1.png}
%    \end{subfigure}%
    \caption{Left to right: ANOVA results for (a) Number of Resets, (b) Distance traveled in PTS, and (c) Total Time Taken. Confidence Interval = 95\%}
    \label{fig:ANOVA_analysis}
\end{figure}

\subsection{Analysis of Results}
A one-way between groups ANOVA $(\alpha = 0.05)$ with repeated measures was performed to compare the effects of with- and without-using the redirection on the dependent variables: (a) number of resets, (b) distance traveled in PTS, (c) total time taken, and (d) scores. We used partial eta squared $(\eta^{2}_{p})$ to report the obtained effect sizes for each variable.

Based on the results of Levene's test \cite{levenestest1960}, it was found that the outcomes for the number of resets ($F (2.14) = 375.710, p > 0.05$), distance traveled ($F (1.176) = 0.348, p > 0.05$) and total time taken ($F (0.971) = 1.001, p > 0.05$) were normally distributed and hence equal variances were assumed. However, the outcome for scores ($F (4.103) = 0.054, p < 0.05$) showed the opposite. As scores violated the homogeneity of variances assumption, the variable was omitted during the ANOVA analysis. 

The results from ANOVA showed a statistically significant difference between the number of resets when the game was played with- and without- using the proposed redirection technique ($F (1, 48) = 375.710, p < 0.001$) with $\eta^{2}_{p} = 0.887$. Nonetheless, these results also showed a statistically insignificant effect of redirection on distance traveled ($F (1, 48) = 0.384, p > 0.05$; $\eta^{2}_{p} = 0.008$), and total time taken ($F (1, 48) = 1.001, p > 0.05$; $\eta^{2}_{p} = 0.020$). The $\eta^{2}_{p}$ values shows that 88.7\% of the variability in the required number of resets is accounted for by our independent variable i.e.- redirection. However, the effects on distance traveled and total time taken remains negligible. The results from this test can be seen in Figure \ref{fig:ANOVA_analysis}. The error bars in the graphs shows a confidence interval of 95\%. 

%The results from ANOVA showed a statistically significant effect of disabling the redirection on the number of resets required to complete the same objective, $F(1, 48) = 375.710, p < 0.001, \eta^{2}_{p} = 0.887$.
Besides this, results of the first experiment also showed that the system applied an average of $1547.55^{\circ}$ (SD = $152.26^{\circ}$) of absolute angular gain to each participant's orientation during the entire test. An average of $3.15^{\circ}$ of absolute angular gain was applied per redirection with an average of 1.73 redirections/s. As the participants were cognitively preoccupied with the task of zapping aliens, they were unaware of this angular gain. Furthermore, since this is a real-time technique and thus the frames were rendered without additional lagging other than what is typically imposed by the hardware, none of the participants reported perceiving any scene nor camera manipulation. In the post-study questionnaire, one of the participant stated that "I felt like I was walking straight. I was completely unaware of my actual movements.".
% Also, 80\% of the participants reported no (TS = 0) to minimal ($TS<10$) signs of simulator sickness in the SSQ.
%\begin{figure}[!ht]
%    \centering
%    \includegraphics[width=0.47\textwidth]{images/Graphs/Resets.png}
%    \caption{Comparison between the number of resets required when the redirection was enabled and disabled}
%    \label{fig:numberOfResets}
%\end{figure}

% \begin{figure}[!ht]
%     \centering
%     \includegraphics[width=0.47\textwidth]{images/FinalSSQ.png}
%     \caption{Results from the SSQ scores (Left to right: Nausea, Oculomotor, Disorientation and Total Severity). The Total Severity and sub-scales such as Nausea, Oculomotor, and Disorientation were calculated based on the formulas in \cite{kennedy1993ssq}}
%     \label{fig:FinalSSQAnalysis}
% \end{figure}

\subsubsection{Simulator Sickness Questionnaire (SSQ)}
After completing the experiment, participants were asked to fill out an SSQ. Based on the scores reported, the majority of the participants (80\%) showed no signs (TS = 0) or minimal signs (TS $<$ 10) of simulator sickness, and only 8\% of the participants reported TS $>$ 12. 
% Similar to the previous user studies (\#1 and \#2), 
The highest score and mean were reported for the disorientation sub-scale even though the rotation angle was at all times well within the limits of tolerance as determined by user study \#1. This can be attributed to the fact that the cognitive workload of the task involved in this user study was more demanding than in the previous user studies. Although this caused an increase in the highest score for disorientation, the mean decreased when compared to that of user study \#2. The median values for all sub-scales, as before, were reported as 0.
% As the user's eyes were exposed to the HMD screens for a longer period of time without any breaks, users temporarily faced difficulty changing their focus right after the experiment. This lead to an increased reported score for difficulty focusing and hence, disorientation. This was caused due to an effect called Computer Vision Syndrome \cite{computerVisionSyndrome}. 
Table \ref{tab:SSQ3} summarizes the results from this SSQ. As it is evident, the mean scores were dropped significantly from user study \#1 and \#2. 

%The increased score of disorientation can be attributed by computer vision syndrome. This was caused as the user's eyes were exposed to the HMD screens for longer periods of time without any break. 

%Users thought that they were walking straight in the VE but actually they were continuously walking in circles in the PTS. This continuous circular motion lead to vertigo and hence the high disorientation score.

%Thus, this can be attributed to the fact that the motion of the peripheral zone induced conflicts between the vestibular and visual signals, leading to vestibular disturbances such as vertigo or dizziness. 

% Figure \ref{fig:FinalSSQAnalysis} shows the scores reported by 25 participants for each sub-scale and the total score for this SSQ.

\begin{table}[!ht]
  \begin{center}
    \caption{Results from the SSQ responses for user study \#3. 80\% of the participants reported no (TS = 0) to minimal ($TS<10$) signs of simulator sickness}
    \label{tab:SSQ3}
    \begin{tabular}{|l|c|c|c|c|c|}
      \hline
      \textbf{Scores} & \textbf{Mean} & \textbf{Median} & \textbf{SD} & \textbf{Min} & \textbf{Max}\\ % <-- added & and content for each column
      \hline
      \textbf{Nausea (N)} & 2.29 & 0 & 5.7 & 0 & 19.08\\ % <--
      \hline
      \textbf{Oculomotor (O)} & 6.67 & 7.58 & 9.1 & 0 & 37.9\\ % <--
      \hline
      \textbf{Disorientation (D)} & \textbf{8.35} & 0 & 17.05 & 0 & \textbf{69.6}\\ % <--
      \hline
      \textbf{Total Score (TS)} & 6.43 & 3.74 & 10.04 & 0 & 44.88\\ % <--
    %   \textbf{N} & 11.27 & 9.54 & 14.67 & 0 & 38.16\\ % <--
    %   \textbf{O} & 19.29 & 15.16 & 24.06 & 0 & 83.38\\ % <--
    %   \textbf{D} & 29.11 & 0 & 45.09 & 0 & 139.2\\ % <--
    %   \textbf{TS} & 21.76 & 7.48 & 28.47 & 0 & 93.5\\ % <--
     \hline
    \end{tabular}
  \end{center}
\end{table}

% %\textbf{I believe if the user walked slow enough, he should be able to walk in long straight paths using their system.}
% %\textbf{They also had this object retrieval task. Find and count the number of predefined objects. Just like the experiment that Christian designed in his multimedia project, last winter.} 
% %\textbf{They did not implement any reset function. They used a smaller walking space then the PTS that they actually had. Hence they are reporting the error area for when the users walked outside the predefined walking space.}

\section{Discussion}
The results of the user study \#3 are indicative of the efficacy of the proposed technique in VR applications where the cognitive workload on the user is moderate. Examples of such applications are immersive games, training simulations, cinematic VR, etc. 

Further to exploiting inattentional blindness, and unlike other state-of-the-art techinques, our technique relies only on \textit{naturally occurring} saccades and blinks, and not stimulated. This is distinct from other reported work in the literature and an important advantage since stimulating saccades is both, disruptive to the cognitive task at hand and increases the effects of VR sickness. For example, in \cite{sun2018towards} saccades are stimulated by introducing orbs of light in image- and object-space to forcefully divert the user's attention in order to perform the redirection. In contrast to this, and based on the psychological effect of inattentional blindness, rather than divert the user's attention we exploit the fact that the user is fixated on a particular task leading to "tunnel-vision"-like focus. This allows us to constantly update in real-time the non-foveal (peripheral) zone without the user perceiving a change. Metaphorically speaking, the foveal vision/zone acts as an update-brush of the framebuffer: whenever it moves, based on the tracking of the user's eyes, everything \textit{within} the foveal zone is rendered from the $Cam_{foveal}$ without any rotations being applied, and everything outside i.e. the non-foveal zone, is rendered from the $Cam_{non-foveal}$ with a rotation $0 < \theta_{Cam_{non-foveal}} < 13.5^{\circ}$ applied to the VE calculated in real-time based on the required redirection.

The experiment in user-study \#3 used a PTS of $4\times4m^{2}$. The results show that even with room-scale PTS such as the one used, the users were able to walk distances in the VE which are up to almost $18\times$ orders of magnitude higher than the longest distance in the PTS (i.e. diagonal of $\sqrt 32$). The maximum recorded distance in our experiments was $103.9m$ with no resets. Furthermore, the traveled distance can include long straight walks as shown in Figure \ref{fig:teaser} and Figure \ref{fig:paths}.

\section{Conclusion and Future Work}
In this work we presented a rotation-based redirection technique using dynamic-foveated rendering which leverages the effect of inattentional blindness induced by a cognitive task. The technique uses natural visual suppressions such as eye blinks and saccades (\textit{without any artificial stimuli}) to make subtle rotations to the VE without the user's knowledge. Furthermore, we conducted extensive tests and presented the results of three user studies. The results confirmed that the technique is indeed effective and can handle long-straight walks. This allows the users to freely explore open world VEs.

Currently, our technique only uses rotational gains for redirection. In the future work, we plan to incorporate translational gains while maintaining the current real-time performance.
% We also plan to incorporate the Dynamic Object in PTS support. 

Lastly, there is an ample amount of research opportunities in enhancing the redirected walking systems which include saccade prediction algorithms and using other forms of visual suppression e.g. phase of nystagmus, etc. 
% Our system works for a 4X4m of PTS. Nonetheless, it would also be interesting to explore the opportunities for the introduction of additional gains in case of bigger PTS. 

\acknowledgments{
This work was supported by the
Natural Sciences and Engineering Research Council of Canada Grants DG-N01670 (Discovery Grant).}

\bibliographystyle{abbrv-doi}

\bibliography{template}

\begin{thebibliography}{10}

\bibitem{mahdi2017toolkit}
M.~Azmandian, T.~Grechkin, and E.~S. Rosenberg.
\newblock An evaluationof strategies for two-user redirected walking in shared
  physical spaces.
\newblock {\em IEEE Virtual Reality}, pp. 91--98, 2017.

\bibitem{bahill1975saccadeSpeedRange}
A.~T. Bahill, M.~R. Clark, and L.~Stark.
\newblock The main sequence, a tool for studying human eye movements.
\newblock {\em Mathematical biosciences}, 24(3-4):191--204, 1975.

\bibitem{bentivoglio1997blinkrate}
A.~R. Bentivoglio, S.~B. Bressman, E.~Cassetta, D.~Carretta, P.~Tonali, and
  A.~Albanese.
\newblock Analysis of blink rate patterns in normal subjects.
\newblock {\em Movement Disorder}, 12:1028--1034, 1997.

\bibitem{benjamin2015saccadeThresholds}
B.~Bolte and M.~Lappe.
\newblock Subliminal reorientation and repositioning in immersive virtual
  environments using saccadic suppression.
\newblock {\em IEEE Transactions of Visualization and Computer Graphics},
  21(4):545--552, 2015.

\bibitem{burr1994saccadeTimings}
D.~C. Burr, M.~C. Morrone, and J.~Ross.
\newblock Selective suppression of the magnocellular visual pathway during
  saccadic eye movements.
\newblock {\em Nature}, 371(6497):511--513, 1994.

\bibitem{cheng2015turkdeck}
L.-P. Cheng, T.~Roumen, H.~Rantzsch, S.~K{\"o}hler, P.~Schmidt, R.~Kovacs,
  J.~Jasper, J.~Kemper, and P.~Baudisch.
\newblock Turkdeck: Physical virtual reality based on people.
\newblock In {\em Proceedings of the 28th Annual ACM Symposium on User
  Interface Software \& Technology}, pp. 417--426. ACM, 2015.

\bibitem{christensen2000inertial}
R.~R. Christensen, J.~M. Hollerbach, Y.~Xu, and S.~G. Meek.
\newblock Inertial-force feedback for the treadport locomotion interface.
\newblock {\em Presence: Teleoperators \& Virtual Environments}, 9(1):1--14,
  2000.

\bibitem{christou2016navigation}
C.~Christou, A.~Tzanavari, K.~Herakleous, and C.~Poullis.
\newblock Navigation in virtual reality: Comparison of gaze-directed and
  pointing motion control.
\newblock In {\em 2016 18th Mediterranean Electrotechnical Conference
  (MELECON)}, pp. 1--6. IEEE, 2016.

\bibitem{darken1997omnitreadmill}
C.~W. Darken, R.~P. and D.~Carmein.
\newblock The omni-directional treadmill: A locomotion device for virtual
  worlds.
\newblock In {\em Proceedings of the 10th Annual ACM Symposium on User
  Interface Software and Technology}, pp. 213--221. ACM, New York, 1997.

\bibitem{dong2017smooth}
Z.-C. Dong, X.-M. Fu, C.~Zhang, K.~Wu, and L.~Liu.
\newblock Smooth assembled mappings for large-scale real walking.
\newblock {\em ACM Transactions on Graphics (TOG)}, 36(6):211, 2017.

\bibitem{fernandes2003cybersphere}
K.~J. Fernandes, V.~Raja, and J.~Eyre.
\newblock Cybersphere: The fully immersive spherical projection system.
\newblock {\em Communications of the ACM}, 46(9):141--146, 2003.

\bibitem{yongHe2014multiRateShading}
Y.~He, Y.~Gu, and K.~Fatahalian.
\newblock Extending the graphics pipeline with adaptive, multi-rate shading.
\newblock {\em ACM Trans. Graph.}, 33(4):142:1--142:12, July 2014. doi: {{%
10\hspace{.1pt}\discretionary{.}{%
}{.}\hspace{.4pt}1145\discretionary{/}{%
}{/}2601097\hspace{.1pt}\discretionary{.}{%
}{.}\hspace{.4pt}2601105}}


\bibitem{hodgson2013SteerToMultipleCenters}
E.~Hodgson and E.~Bachmann.
\newblock Comparing four approaches to generalized redirected walking:
  Simulation and live user data.
\newblock {\em IEEE transactions on visualization and computer graphics},
  19:634--643, 2013.

\bibitem{hodgson2014constrainedVE}
E.~Hodgson, E.~Bachmann, and T.~Thrash.
\newblock Performance of redirected walking algorithms in a constrained virtual
  world.
\newblock {\em IEEE transactions on visualization and computer graphics},
  20:579--587, 2014.

\bibitem{huang2003strollBased}
J.~Y. Huang.
\newblock An omnidirectional stroll-based virtual reality interface and its
  application on overhead crane training.
\newblock {\em IEEE Transactions on Multimedia}, 5(1):39--51, 2003.

\bibitem{ibbotson2009unknownReasoning}
M.~R. Ibbotson and S.~L. Cloherty.
\newblock Visual perception: Saccadic omission- suppression or temporal
  masking?
\newblock {\em Current Biology}, 19(12):R493--R496, 2009.

\bibitem{iwata1999infiniteFloor}
H.~Iwata.
\newblock Walking about virtual environments on an inﬁnite ﬂoor.
\newblock {\em IEEE Virtual Reality}, pp. 286--293, 1999.

\bibitem{iwata1996perambulator}
H.~Iwata and T.~Fujii.
\newblock Virtual perambulator: a novel interface device for locomotion in
  virtual environment.
\newblock In {\em Proceedings of the IEEE 1996 Virtual Reality Annual
  International Symposium}, pp. 60--65. IEEE, 1996.

\bibitem{kennedy2003ssq}
R.~Kennedy, J.~Drexler, D.~Compton, K.~Stanney, D.~Lanham, and D.~Harm.
\newblock Configural scoring of simulator sickness, cybersickness and space
  adaptation syndrome: Similarities and differences.
\newblock {\em Virtual and Adaptive Environments: Applications, implications
  and human performance issues}, p. 247, 2003.

\bibitem{kennedy1993ssq}
R.~S. Kennedy, N.~E. Lane, K.~S. Berbaum, and M.~G. Lilienthal.
\newblock Simulator sickness questionnaire: An enhanced method for quantifying
  simulator sickness.
\newblock {\em The International Journal of Aviation Psychology}, 3:203--220,
  1993.

\bibitem{langbehn2016blinksThreshold}
E.~Langbehn, G.~Bruder, and F.~Steinicke.
\newblock Subliminal re-orientation and re-positioning in virtual reality
  during eye blinks.
\newblock pp. 213--213, 2016.

\bibitem{langbehn2017predefinedCurvedPaths}
E.~Langbehn, P.~Lubos, G.~Bruder, and F.~Steinicke.
\newblock Application of redirected walking in room-scale vr. in virtual
  reality.
\newblock {\em IEEE Virtual Reality}, 2017.

\bibitem{langbehn2018blinks}
E.~Langbehn and F.~Steinicke.
\newblock In the blink of an eye - leveraging blink-induced suppression for
  imperceptible position and orientation redirection in virtual reality.
\newblock {\em ACM Transactions on Graphics}, 37:1--11, 2018.

\bibitem{laviola2000cybersickness}
J.~J. LaViola~Jr.
\newblock A discussion of cybersickness in virtual environments.
\newblock {\em ACM SIGCHI}, 32(1):47--56, 2000.

\bibitem{levenestest1960}
H.~Levene.
\newblock Robust testes for equality of variances.
\newblock {\em Contributions to Probability and Statistics}, pp. 278--292,
  1960.

\bibitem{medina2008virtusphere}
E.~Medina, R.~Fruland, and S.~Weghorst.
\newblock Virtusphere: Walking in a human size vr “hamster ball”.
\newblock In {\em Proceedings of the Human Factors and Ergonomics Society
  Annual Meeting}, vol.~52, pp. 2102--2106. SAGE Publications Sage CA: Los
  Angeles, CA, 2008.

\bibitem{1999_moshell}
M.~Moshell.
\newblock Infinite virtual walk.
\newblock In {\em Personal Communication}, 1999.

\bibitem{nagamori2005ballArrayTreadmill}
A.~Nagamori, K.~Wakabayashi, and M.~Ito.
\newblock The ball array treadmill: A locomotion interface for virtual worlds.
\newblock {\em IEEE Virtual Reality}, pp. 3--6, 2005.

\bibitem{nvidiaVRS}
{NVIDIA}.
\newblock {VRWorks - Variable Rate Shading}.
\newblock {https://developer.nvidia.com/vrworks/graphics/variablerateshading}
  (last accessed 4th Nov. 2019), 2018.

\bibitem{zerolightVRS}
{O'Connor, Chris}.
\newblock {ZeroLight Improves Automotive Product Visualisation Quality and
  Performance with VRS}.
\newblock {https://developer.nvidia.com/vrworks/graphics/variablerateshading}
  (last accessed 4th Nov. 2019), 2018.

\bibitem{regan2000temporaryBlindness}
J.~K. ORegan, H.~Deubel, J.~J. Clark, and R.~A. Rensink.
\newblock Picture changes during blinks: Looking without seeing and seeing
  without looking.
\newblock {\em Visual cognition}, 7:191--211, 2000.

\bibitem{patney2016perceptuallyBasedFR}
A.~Patney, J.~Kim, M.~Salvi, A.~Kaplanyan, C.~Wyman, N.~Benty, A.~Lefohn, and
  D.~Luebke.
\newblock Perceptually-based foveated virtual reality.
\newblock In {\em ACM SIGGRAPH 2016 Emerging Technologies}, SIGGRAPH '16, pp.
  17:1--17:2. ACM, New York, NY, USA, 2016. doi: {{%
10\hspace{.1pt}\discretionary{.}{%
}{.}\hspace{.4pt}1145\discretionary{/}{%
}{/}2929464\hspace{.1pt}\discretionary{.}{%
}{.}\hspace{.4pt}2929472}}


\bibitem{peck2011walkInPlace}
T.~Peck, H.~Fuchs, and M.~Whitton.
\newblock An evaluation of navigational ability comparing redirected free
  exploration with distractors to walking-in-place and joystick locomotion
  interfaces.
\newblock pp. 56--62. IEEE, 2011.

\bibitem{poullis2009automatic}
C.~Poullis and S.~You.
\newblock Automatic creation of massive virtual cities.
\newblock In {\em 2009 IEEE Virtual Reality Conference}, pp. 199--202. IEEE,
  2009.

\bibitem{ramot2008blinkDuration}
{Ramot, Daniel}.
\newblock {Average duration of a single blink}.
\newblock {https://bionumbers.hms.harvard.edu/bionumber.aspx?id=100706\&ver=0}
  (last accessed 1st Nov. 2019).

\bibitem{razzaques2001redirected}
S.~Razzaque, Z.~Kohn, and M.~C. Whitton.
\newblock Redirected walking.
\newblock {\em Proceedings of Eurographics}, 9:105--106, 2001.

\bibitem{razzaque2005redirected}
S.~Razzaque, Z.~Kohn, and M.~C. Whitton.
\newblock {\em Redirected walking}.
\newblock Citeseer, 2005.

\bibitem{razzaques2002redirected}
S.~Razzaque, D.~Swapp, M.~Slater, M.~C. Whitton, and A.~Steed.
\newblock Redirected walking in place.
\newblock {\em Eurographics Symposium on Virtual Environments}, pp. 123--130,
  2002.

\bibitem{reder1973FirstFoveatedRender}
S.~M. Reder.
\newblock On-line monitoring of eye-position signals in contingent and
  noncontingent paradigms.
\newblock {\em Behavior Research Methods \& Instrumentation}, 5:218--228, 1973.

\bibitem{rensink2002temporaryBlindness}
R.~A. Rensink.
\newblock Change detection.
\newblock {\em Annual review of psychology}, 53:245--277, 2002.

\bibitem{rensink1997temporaryBlindness}
R.~A. Rensink, J.~K. O’Regan, and J.~J. Clark.
\newblock To see or not to see:the need for attention to perceive changes in
  scenes.
\newblock {\em Psychological science}, 8:368--373, 1997.

\bibitem{ruddle2009cognitiveMapping}
R.~Ruddle and S.~Lessels.
\newblock Walking interface to navigate virtual environments.
\newblock {\em ACM Transactions on Computer-Human Interaction}, 16:5:1--5:18,
  2009.

\bibitem{ruddle2011cognitiveMapping}
R.~A. Ruddle, E.~P. Volkova, and H.~H. Bulthoff.
\newblock Walking improves your cognitive map in environments that are
  large-scale and large in extent.
\newblock {\em ACM Transactions on Computer-Human Interaction}, 18:10:1--10:22,
  2011.

\bibitem{simons1999gorillas}
D.~J. Simons and C.~F. Chabris.
\newblock Gorillas in our midst: Sustained inattentional blindness for dynamic
  events.
\newblock {\em perception}, 28(9):1059--1074, 1999.

\bibitem{souman2010making}
J.~L. Souman, P.~R. Giordano, I.~Frissen, A.~D. Luca, and M.~O. Ernst.
\newblock Making virtual walking real: Perceptual evaluation of a new treadmill
  control algorithm.
\newblock {\em ACM Transactions on Applied Perception (TAP)}, 7(2):11, 2010.

\bibitem{steinicke2008headRotations}
F.~Steinicke, G.~Bruder, L.~Kohli, J.~Jerald, and K.~Hinrichs.
\newblock Taxonomy and implementation of redirection techniques for ubiquitous
  passive haptic feedback.
\newblock pp. 217--223. IEEE, 2008.

\bibitem{stengel2016peceptuallyGuidedFR}
M.~Stengel, S.~Grogorick, M.~Eisemann, and M.~Magnor.
\newblock {Adaptive Image-Space Sampling for Gaze-Contingent Real-time
  Rendering}.
\newblock {\em Computer Graphics Forum}, 2016. doi: {{%
10\hspace{.1pt}\discretionary{.}{%
}{.}\hspace{.4pt}1111\discretionary{/}{%
}{/}cgf\hspace{.1pt}\discretionary{.}{%
}{.}\hspace{.4pt}12956}}


\bibitem{suma2011leveraging}
E.~A. Suma, S.~Clark, D.~Krum, S.~Finkelstein, M.~Bolas, and Z.~Warte.
\newblock Leveraging change blindness for redirection in virtual environments.
\newblock In {\em 2011 IEEE Virtual Reality Conference}, pp. 159--166. IEEE,
  2011.

\bibitem{suma2012impossible}
E.~A. Suma, Z.~Lipps, S.~Finkelstein, D.~M. Krum, and M.~Bolas.
\newblock Impossible spaces: Maximizing natural walking in virtual environments
  with self-overlapping architecture.
\newblock {\em IEEE Transactions on Visualization and Computer Graphics},
  18(4):555--564, 2012.

\bibitem{sun2018towards}
Q.~Sun, A.~Patney, L.-Y. Wei, O.~Shapira, J.~Lu, P.~Asente, S.~Zhu, M.~Mcguire,
  D.~Luebke, and A.~Kaufman.
\newblock Towards virtual reality infinite walking: dynamic saccadic
  redirection.
\newblock {\em ACM Transactions on Graphics (TOG)}, 37(4):67, 2018.

\bibitem{sun2016mapping}
Q.~Sun, L.-Y. Wei, and A.~Kaufman.
\newblock Mapping virtual and physical reality.
\newblock {\em ACM Transactions on Graphics (TOG)}, 35(4):64, 2016.

\bibitem{usoh1999flying}
M.~Usoh, K.~Arthur, M.~C. Whitton, R.~Bastos, A.~Steed, M.~Slater, and
  J.~Frederick P.~Brooks.
\newblock Walking > walking-in-place > flying, in virtual environments.
\newblock pp. 359--364. ACM, 1999.

\bibitem{volkmann1986suppressions}
F.~C. Volkmann.
\newblock Human visual suppressions.
\newblock {\em Vision Research}, 26:1401--1416, 1986.

\bibitem{volkmann1980blinks}
F.~C. Volkmann, L.~A. Riggs, and R.~K. Moore.
\newblock Eyeblinks and visual suppression.
\newblock {\em Science}, 207(4433):900--902, 1980.

\bibitem{benjamin2013suspendedWalking}
B.~Walther-Franks, D.~Wenig, J.~Smeddinck, and R.~Malaka.
\newblock Suspended walking: A physical locomotion interface for virtual
  reality.
\newblock In {\em International Conference on Entertainment Computing}, pp.
  185--188. Springer Berlin Heidelberg, 2013.

\bibitem{williams2007resetTechniques}
B.~Williams, G.~Narasimham, B.~Rump, T.~P. McNamara, T.~H. Carr, J.~Rieser, and
  B.~Bodenheimer.
\newblock Exploring large virtual environments with an hmd when physical space
  is limited.
\newblock pp. 41--48. ACM, 2007.

\bibitem{zank2015pastWalkingDirection}
M.~Zank and A.~Kun.
\newblock Using locomotion models for estimating walking targets in immersive
  virtual environments.
\newblock IEEE, 2015.

\bibitem{zank2015eyeTracking}
M.~Zank and A.~Kun.
\newblock Eye tracking for locomotion prediction in redirected walking.
\newblock IEEE, 2016.

\end{thebibliography}

\end{document}